\documentclass{PoS}
\usepackage{amsmath,bm}
\usepackage{subfigure}
\usepackage{graphicx}
\usepackage{epsfig}
\graphicspath{{figures/}} 

\title{Radial profile of heavy quarks in jets in high-energy nuclear collisions}

\ShortTitle{Radial profile of heavy quarks in jets in high-energy nuclear collisions}

\author{\speaker{Sa Wang}$^a$, Wei Dai$^b$, Ben-Wei Zhang$^{a,c}$ , Enke Wang$^{c,a}$\\
        \llap{$^a$} Key Laboratory of Quark \& Lepton Physics (MOE) and Institute of Particle Physics, Central China Normal University, Wuhan 430079, China\\
        \llap{$^b$}School of Mathematics and Physics, China University of Geosciences, Wuhan 430074, China\\
         \llap{$^c$}Institute of Quantum Matter, South China Normal University, Guangzhou 510006, China
        E-mail: \email{wangsa@mails.ccnu.edu.cn}\\
        \qquad\qquad\email{bwzhang@mail.ccnu.edu.cn}}

\abstract{
In high energy nuclear collisions, heavy flavor tagged jets are useful hard probes to study the properties of the quark-gluon plasma (QGP). In this talk, we present the first theoretical prediction of the $D^0$ meson radial distributions in jets relative to the jet axis both in p+p and Pb+Pb collisions at $5.02$~TeV, it shows a nice agreement with the available experimental data. The in-medium jet evolution in the study is described by a Monte Carlo transport model which has been incorporated with the initial events as input provided by the next-to-leading order (NLO) plus parton shower (PS) event generator SHERPA. In such evolution process, both elastic and inelastic parton energy loss in the hot and dense medium are taken into account. Within this same simulation framework, we predict different modification patterns of the radial profile of charm and bottom quarks in jets in Pb+Pb collisions: jet quenching effect will lead the charm quarks diffuse to lager radius while lead the bottom quarks distributed closer to jet axis.

}

\FullConference{International Conference on Hard and Electromagnetic Probes of High-Energy Nuclear Collisions\\
		31 May 2020 to 5 June 2020\\
		Online}

\begin{document}

\section{Introduction}
\label{Intro}
In high energy heavy-ion collisions at the Relativistic Heavy Ion Collider (RHIC) and the Large Hadron Collider (LHC) experiments, the de-confined state of quark-gluon plasma (QGP) formed under such extreme condition provides an arena to manifest the interaction properties of Quantum Chromodynamics (QCD)  matter. The strong interaction between the high-$p_T$ jet produced in the hard scattering with the hot and dense medium created later, referred as jet quenching effect~\cite{Vitev:2009rd}, has been extensively investigated to study the properties of the QGP. Especially, owing to the early creation time, heavy quarks can survive when traverse through the bulk medium, hence the heavy quark tagged jets are useful hard probes to study the properties of the QGP.

Recently, the $D^0$ meson radial distributions in jets relative to the jet axis both in p+p and Pb+Pb collisions measured by CMS collaboration~\cite{Sirunyan:2019dow} reveal a significant diffusion behavior of charm quark in jets due to the in-medium interaction, it provide a new aspect to study the mass effect of jet quenching. In this talk, we present the first theoretical prediction of $D^0$ radial profile in jets in heavy-ion collisions~\cite{Wang:2019xey,Wang:2020bqz}. To figure out the mass dependence of the radial profile in-medium modification of heavy quarks in jets, we compare the radial distributions of the charm and the bottom quarks in jets in p+p collisions and study their medium modification in Pb+Pb collisions based on it.

\section{p+p baseline and in-medium parton energy loss}
\label{baseline}

\begin{figure}[!t]
\begin{center}
\vspace*{-0.1in}
\hspace*{-.1in}
\subfigure[]{
  \epsfig{file=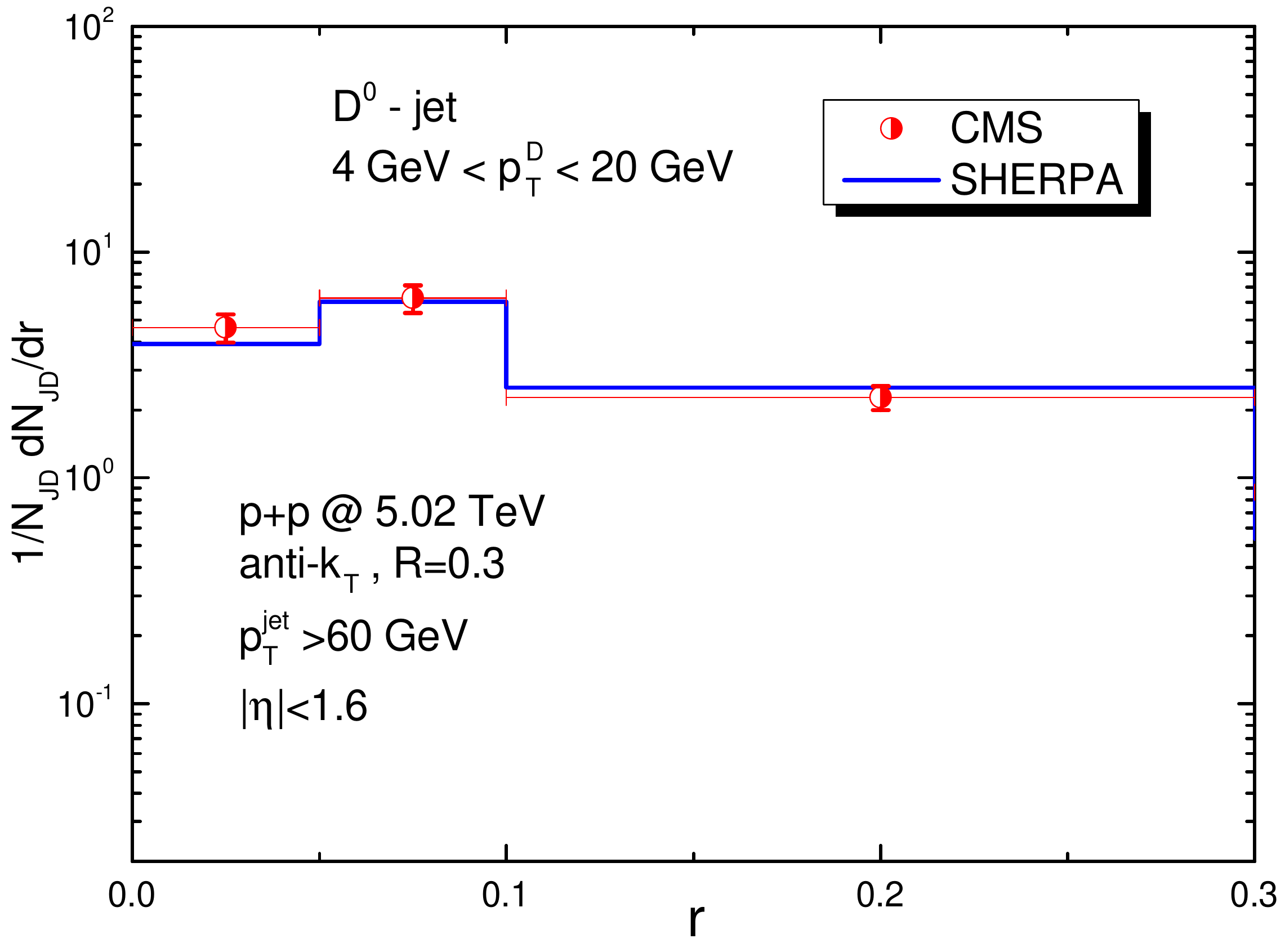, width=2.5in,height=2.3in,angle=0, clip=}}
 \subfigure[]{
  \epsfig{file=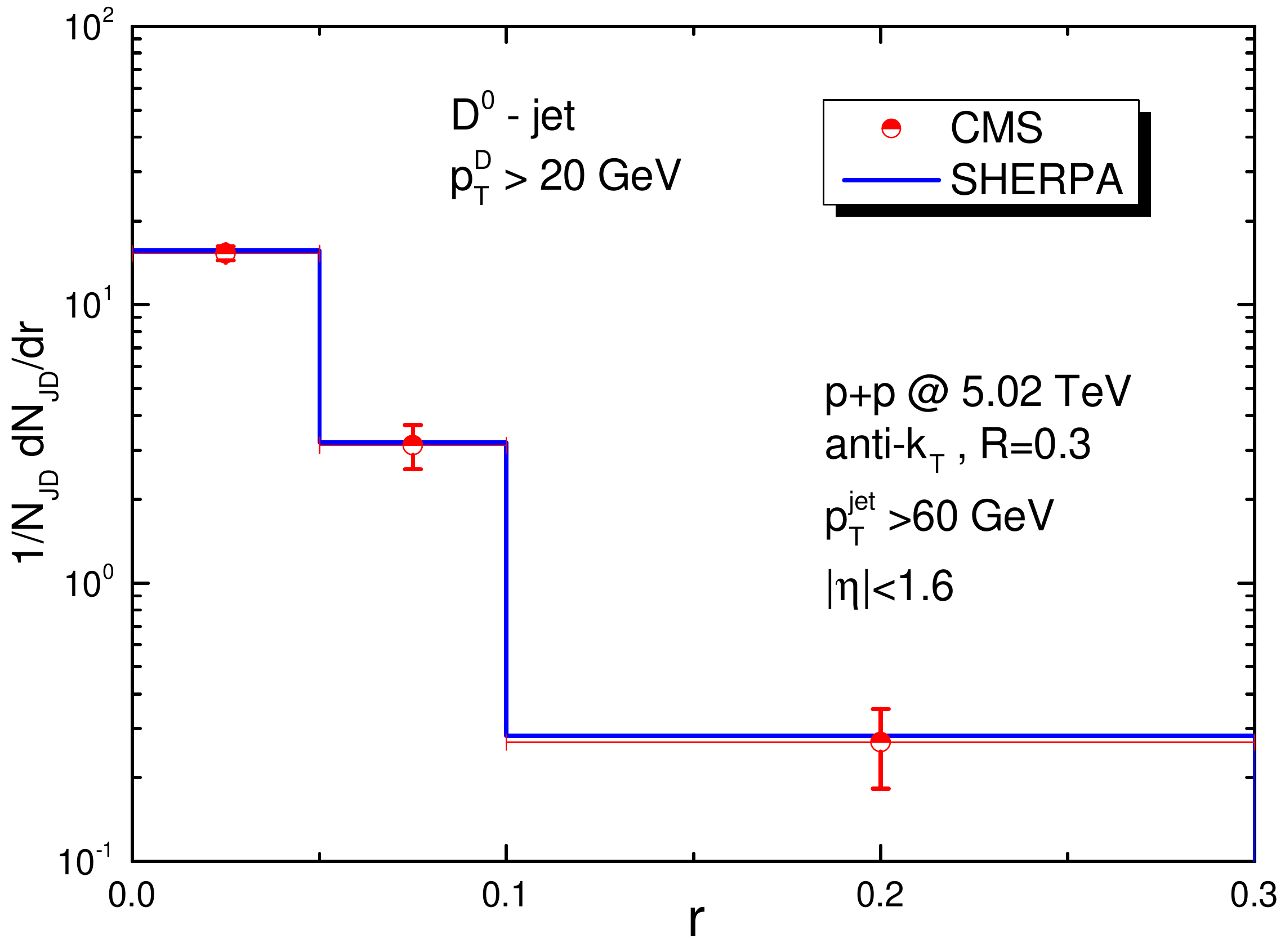,width=2.5in,height=2.3in, clip=}}
\vspace*{-.2in}
\hspace*{0.2in}
\caption{(Color online) The normalized radial distributions of $D^0$ meson in jets as a function of the angular distance from the jet axis in p+p collisions at $5.02$~TeV provided by SHERPA are compared with CMS data at (a) $4$~GeV $< p^D_T < 20$~GeV and (b) $p^D_T > 20$~GeV.}
\label{fig:dndrpp}
\end{center}
\end{figure}

 In this work, we use a Monte Carlo event generator SHERPA~\cite{Gleisberg:2008ta}, which matches the next-to-leading order (NLO) QCD matrix elements with parton shower (PS), to provide the p+p baseline. Due to the too low possibility of coalescence at $p_{HQ}>4$~GeV, we perform the hadronization of heavy quark by the Peterson form fragmentation functions (FFs) ~\cite{Peterson:1982ak}. And we assume that light parton hadronization would not significantly change the full jet momentum as well as the radial distribution of $D^0$ mesons in jet. A decent description on the experimental data is found by the simulation of SHERPA shown in Fig.~\ref{fig:dndrpp} for two kinematic regions, where $r=\sqrt{(\Delta\phi_{JD})^2+(\Delta\eta_{JD})^2}$ is defined by the relative azimuthal angle $\Delta\phi_{JD}$ and relative pseudorapidity $\Delta\eta_{JD}$ between the $D^0$ meson and the jet axis, all the selected jets are reconstructed by anti-$k_T$ algorithm with $R=0.3$ within $|\eta^{\rm jet}|<1.6$~\cite{Cacciari:2011ma}.

To describe the in-medium heavy flavor jet evolution, elastic and inelastic partonic interactions are both considered. We use the modified Langevin equations incorporating with a recoil term to describe the heavy quark propagation~\cite{Cao:2013ita,Dai:2018mhw,Wang:2020qwe},
\begin{eqnarray}
&\vec{x}(t+\Delta t)&=\vec{x}(t)+\frac{\vec{p}(t)}{E}\Delta t \\
&\vec{p}(t+\Delta t)&=\vec{p}(t)-\Gamma(p)\vec{p} \Delta t+\vec{\xi}(t)-\vec{p}_{\rm g}
\label{eq:lang}
\end{eqnarray}
where drag coefficient $\Gamma$ and diffusion coefficient $\kappa$ are related by the so called fluctuation-dissipation relation $\kappa=2ET\Gamma$ and their values are determined by the Lattice QCD calculation~\cite{Francis:2015daa}. The medium-induced gluon radiation was sampled based on the higher-twist approach~\cite{Guo:2000nz,Zhang:2003yn,Zhang:2003wk,Majumder:2009ge}:
\begin{eqnarray}
\frac{dN}{dxdk^{2}_{\perp}dt}=\frac{2\alpha_{s}C_sP(x)\hat{q}}{\pi k^{4}_{\perp}}\sin^2(\frac{t-t_i}{2\tau_f})(\frac{k^2_{\perp}}{k^2_{\perp}+x^2M^2})^4
\end{eqnarray}
 where $x$ and $k_\perp$ are the energy fraction and transverse momentum carried by the radiated gluon, the last quadruplicate term denote dead-cone effect of massive heavy quarks. $\hat{q}$ is the jet transport parameter extracted from a global fitting of single hadron production in A+A collisions~\cite{Ma:2018swx}. Since Langevin treatment is not suitable for the massless light partons, we used the results of pQCD calculation at Hard Thermal Loop~(HTL) approximation~\cite{Neufeld:2010xi} to take into account their collisional energy loss. The space-time evolution of the expanding QGP medium is provided by the VISHNEW hydrodynamic model~\cite{Shen:2014vra}.

\section{Results and discussions}
\label{sec:results}

\begin{figure}[!t]
\begin{center}
\vspace*{-0.1in}
\hspace*{-.1in}
\subfigure[]{
  \epsfig{file=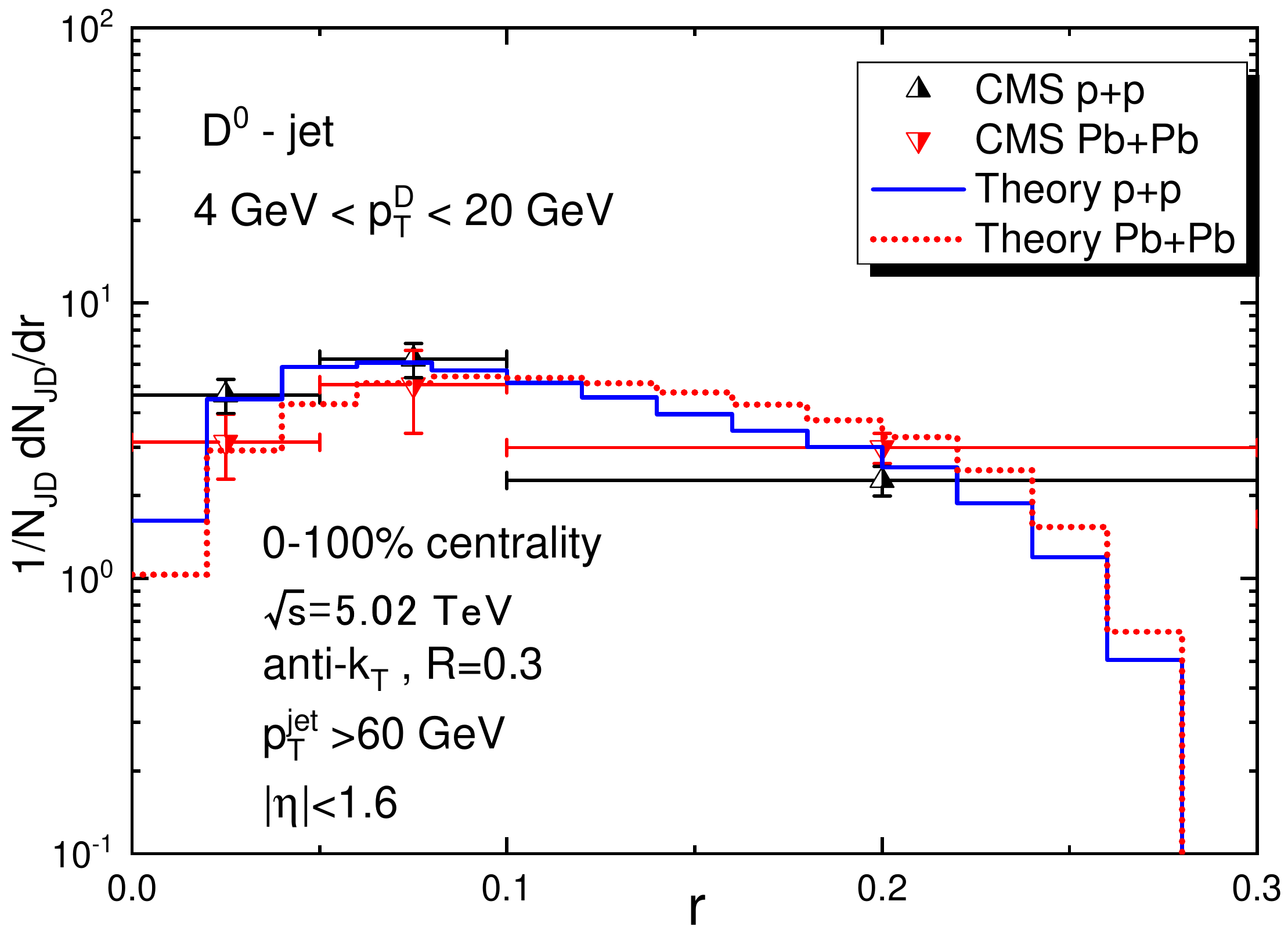, width=2.5in,height=2.3in,angle=0, clip=}}
 \subfigure[]{
  \epsfig{file=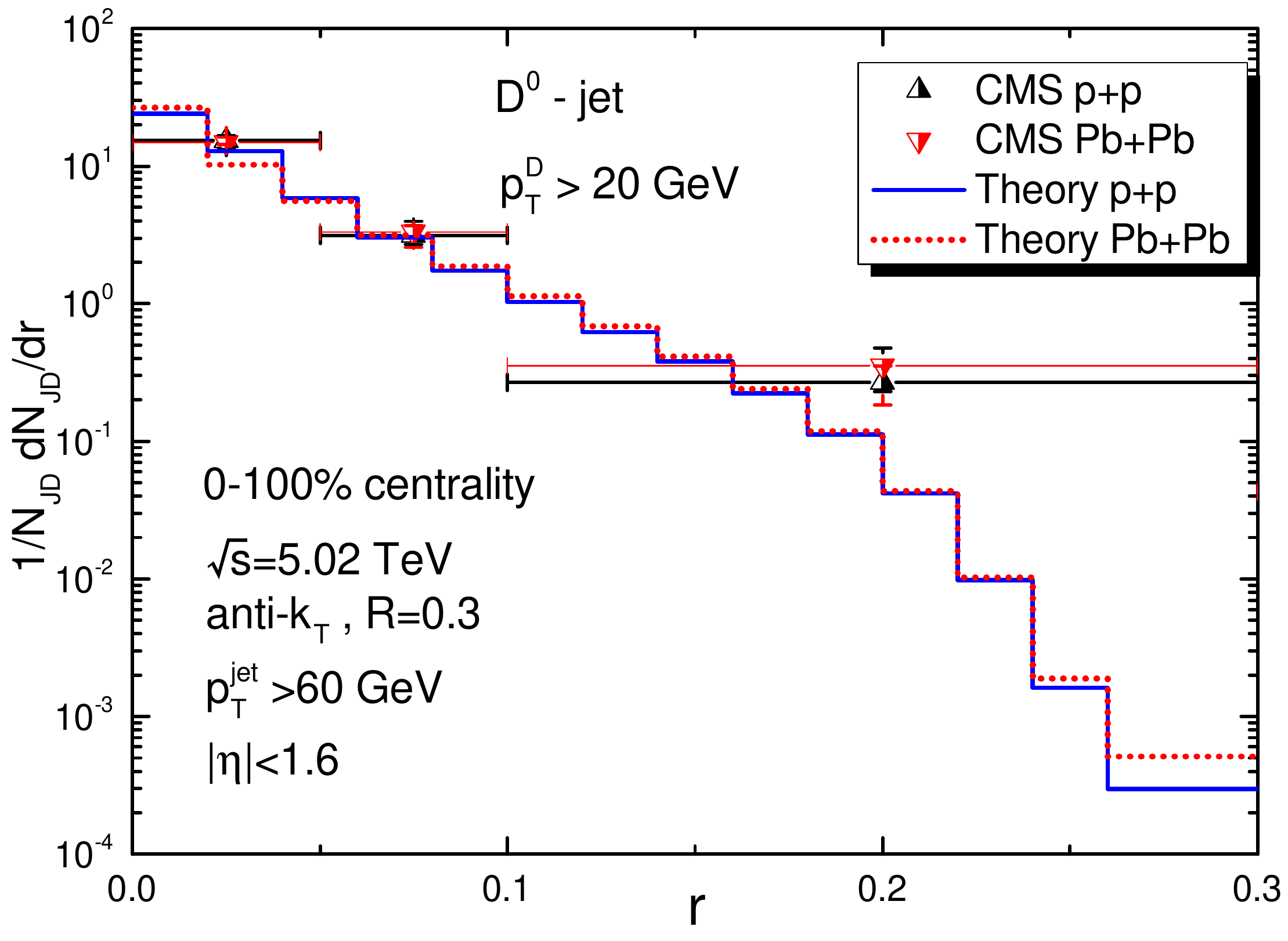,width=2.5in,height=2.3in,angle=0, clip=}}
\vspace*{-.1in}
\hspace*{.2in}
\caption{(Color online) Radial distribution of $D^0$ meson in jets in p+p and Pb+Pb collisions at two $p_T$ range: (a) $4$~GeV $< p^D_T < 20$~GeV and (b) $p^D_T > 20$~GeV are compared with the CMS measurements~\cite{Sirunyan:2019dow}.}
\label{fig:ratio}
\end{center}
\end{figure}

As shown in Fig.~\ref{fig:ratio}, to understand the results measured by CMS~\cite{Sirunyan:2019dow}, we calculate the radial profile of $D^0$ meson in jets at two kinematic regions( $4$~GeV$<p^D_T<20$~GeV and $p^D_T>20$~GeV) both in p+p and Pb+Pb collisions at 5.02 TeV. At $4<p^D_T<20$~GeV, suppression near jet axis and enhancement at larger radius indicates that charm quarks diffuse farther away from the jet axis due to the in-medium interaction in QGP. Whereas, at $p^D_T>20$~GeV, the modification is too small. This diffusion effect can be interpreted by two mechanisms: the random kicks by medium constituents in Brown motion and the momentum recoil by medium-induced gluon radiation, especially at lower charm quark $p_T$~($<$5~GeV) the former plays a critical role~\cite{Wang:2019xey,Wang:2020bqz}.

\begin{figure}[!t]
\begin{center}
\vspace*{-0.1in}
\hspace*{-.1in}
\subfigure[]{
  \epsfig{file=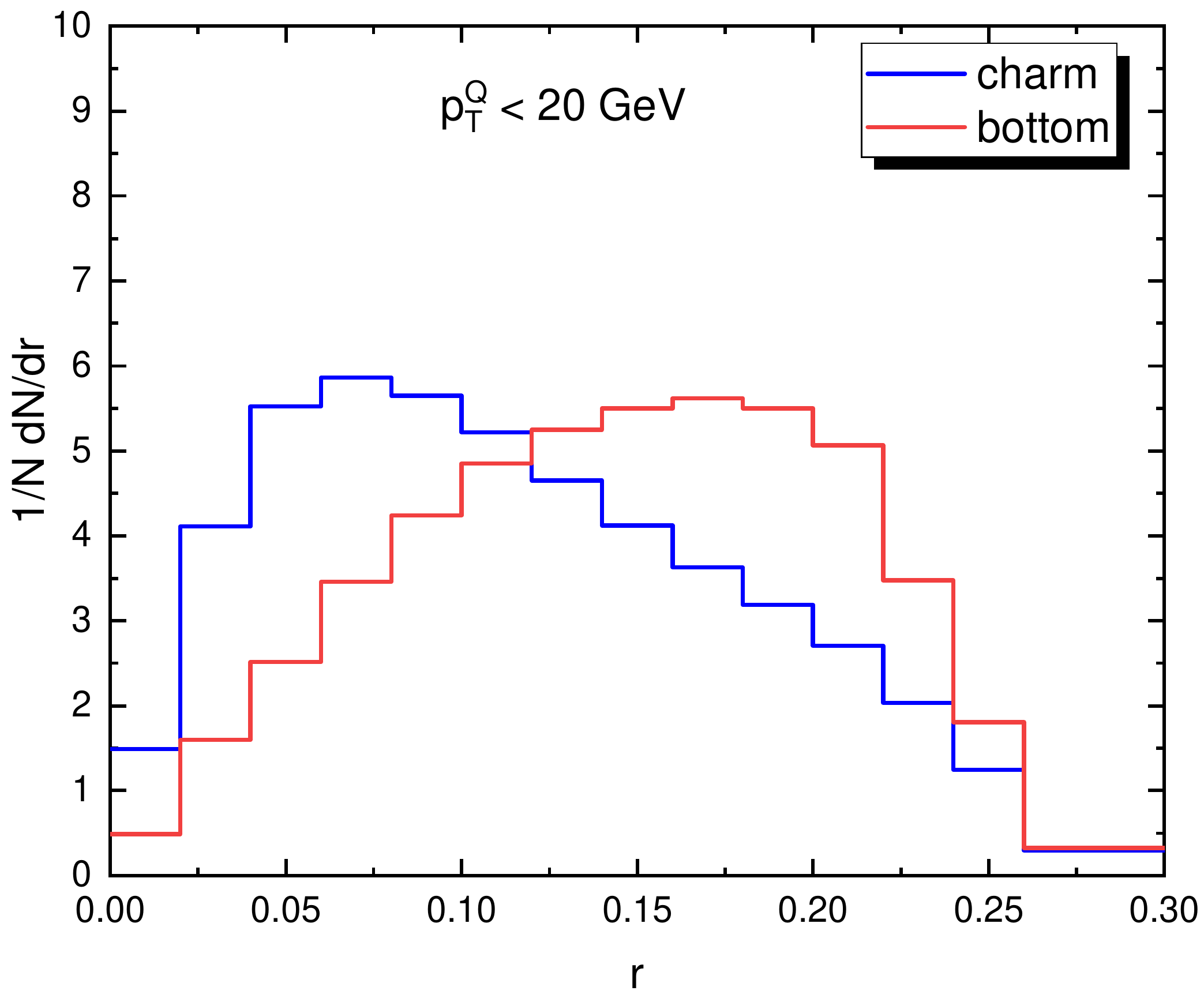, width=1.9in,height=1.7in,angle=0, clip=}}
\subfigure[]{
  \epsfig{file=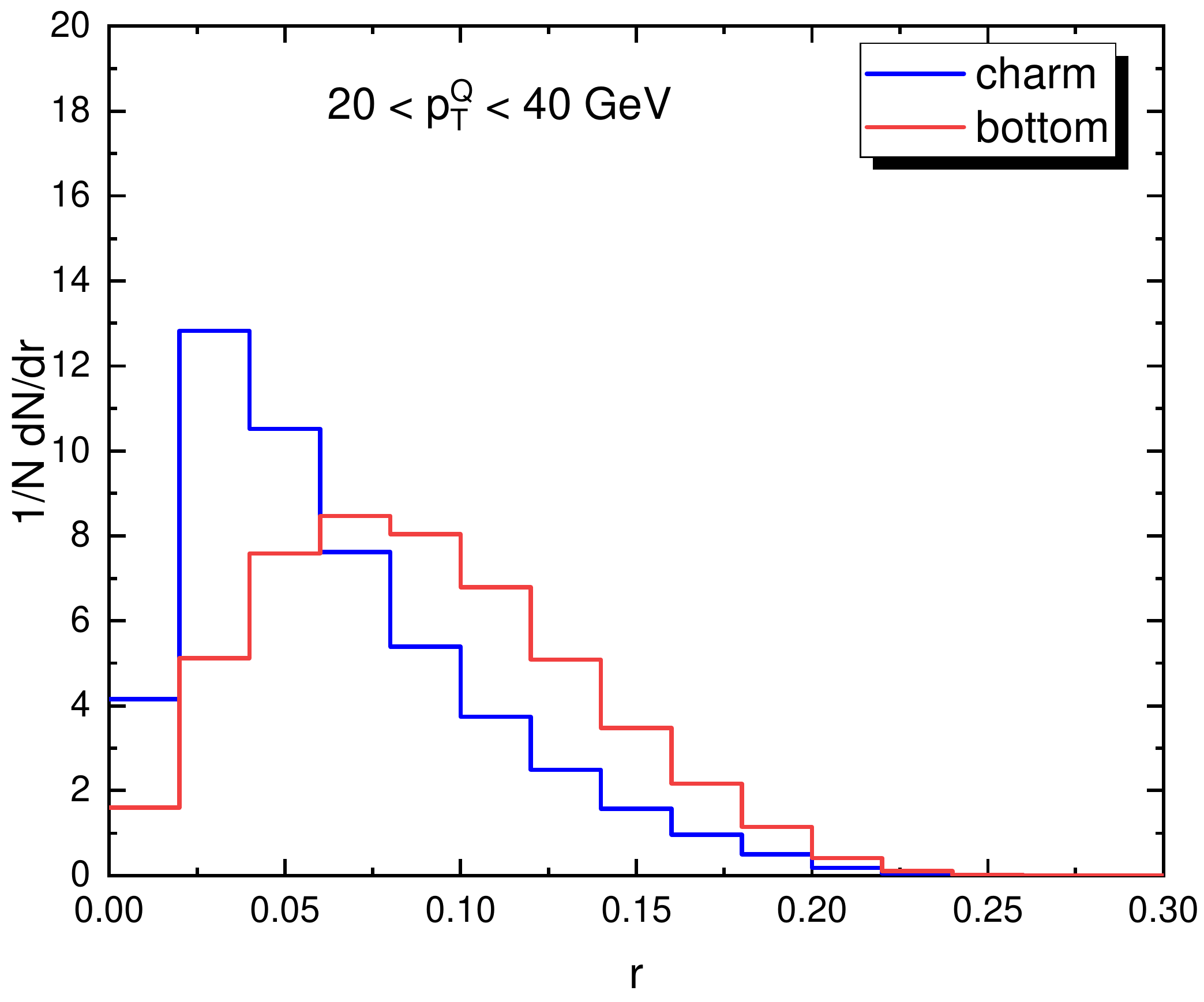, width=1.9in,height=1.7in,angle=0, clip=}}
\subfigure[]{
  \epsfig{file=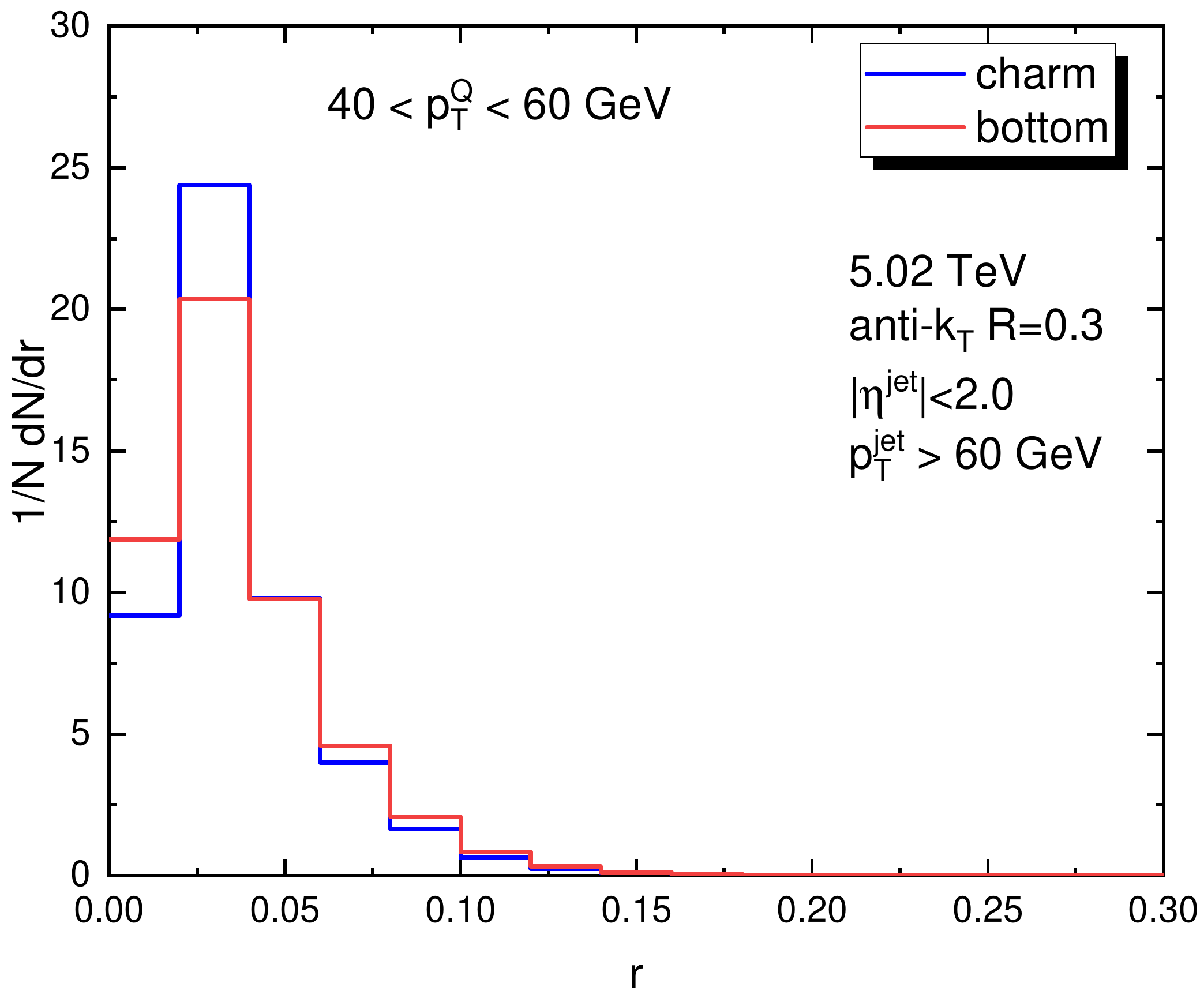,width=1.9in,height=1.7in,angle=0, clip=}}
\vspace*{-.1in}
\hspace*{.2in}
\caption{(Color online) Radial distributions of charm and bottom quarks in jets in p+p collisions for three $p^Q_T$ ranges: (a) $p^Q_T < 20$~GeV, (b) 20 GeV $< p^Q_T < 40$~GeV, and (c) 40 GeV $< p^Q_T < 60$~GeV.}
\label{fig:radial-bc-pp}
\end{center}
\end{figure}

\begin{figure}[!t]
\begin{center}
\vspace*{-0.1in}
\hspace*{.2in}
\subfigure[]{
  \epsfig{file=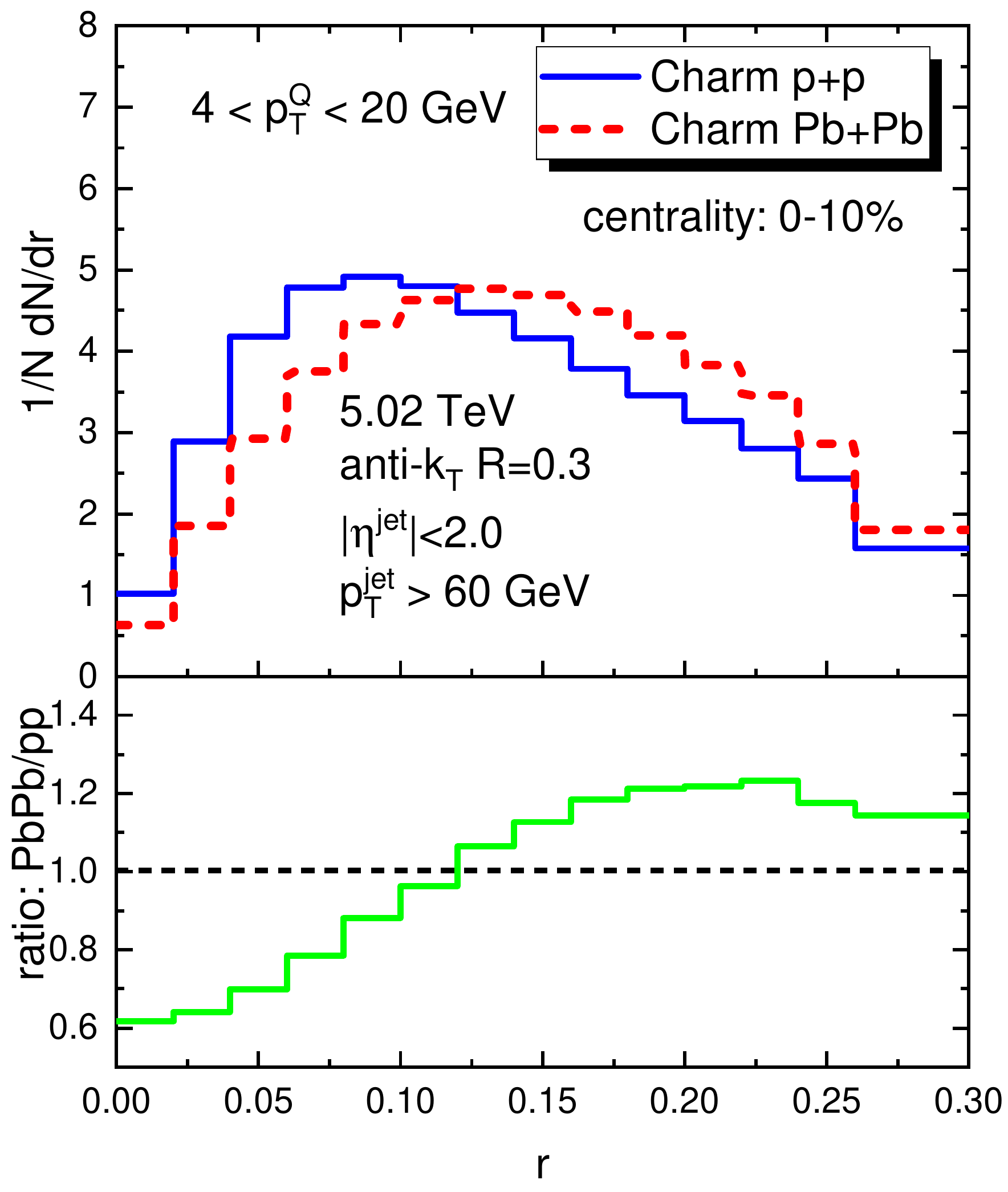, width=2.5in,height=3in,angle=0, clip=}}
 \subfigure[]{
  \epsfig{file=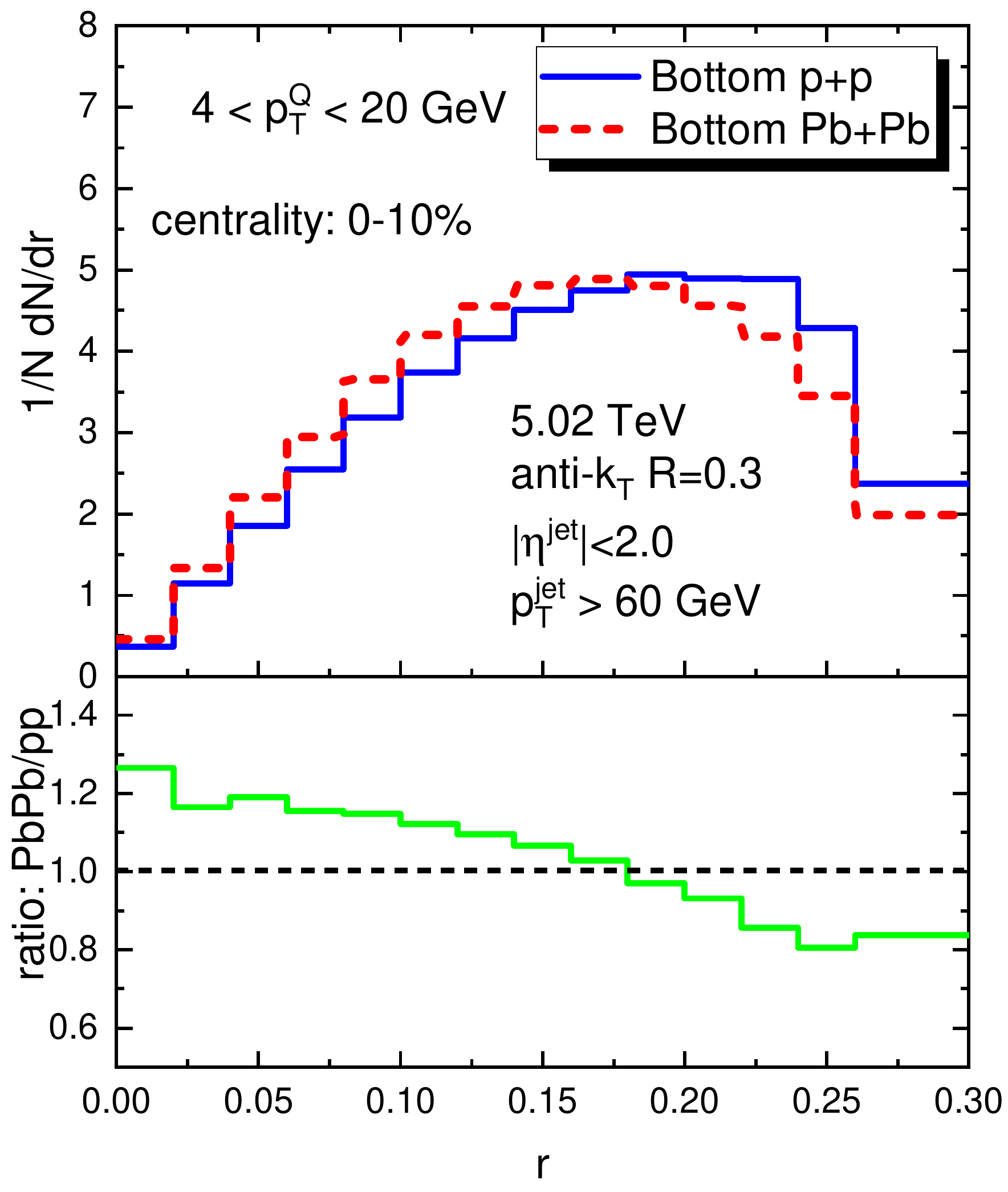,width=2.5in,height=3in,angle=0, clip=}}
\vspace*{-.1in}
\hspace*{.2in}
\caption{(Color online) Radial distributions of charm (a) and bottom (b) quarks in jets in p+p and $0-10\%$ Pb+Pb collisions at $\sqrt{s_{NN}}=5.02$ TeV, for the $p^Q_T$ range : 4 GeV $< p^Q_T < 20$~GeV, the ratios (PbPb/pp) have also been shown in the lower panels to quantify the medium modification. }
\label{fig:radial-cb-aa}
\end{center}
\end{figure}
To figure out the mass effects of charm and bottom quarks of their radial profiles in jets, in Fig.~\ref{fig:radial-bc-pp}, we compare the radial distributions of charm and bottom quarks in jets in p+p collisions at 5.02~TeV at three kinematic regions: $0-20$~GeV, $20-40$~GeV, $40-60$~GeV. With the same configuration, at $p_T^Q<20$~GeV, we find that bottom quarks are farther apart from the jet axis than charm quarks, and the peak of bottom radial distribution is found at $r=0.15$ while charm at $r=0.05$. When it goes to higher $p_T$, the distributions shift to near the jet axis, and the distinctions between charm and bottom trend to disappear.
In Fig.\ref{fig:radial-cb-aa}, we present the radial profile of charm (left) and bottom (right) quarks in jets in p+p and central $0-10\%$ Pb+Pb collisions at $\sqrt{s_{NN}}=$~5.02 TeV, as well as the ratios (PbPb/pp) denoting the medium modifications. We observe two distinct medium modification patterns shown in the ratios: charm quarks shift towards lager radius while bottom quarks seems shift closer to the jet axis. And we confirm that the different modification patterns shown in the radial profile of charm and bottom jets are mainly caused by their different initial distributions~\cite{Wang:2020gxz}.

In summary, we present the first theoretical calculation of the heavy quark radial distribution in jets in heavy-ion collisions to study the diffusion mechanisms of charm quarks in QGP. By comparing the radial profile of charm and bottom in jets both in p+p and Pb+Pb, we also find different modification patterns between bottom jet and charm jet, jet quenching effect will lead the charm quarks diffuse to lager radius while lead the bottom quarks distributed closer to jet axis.

This research is supported by the NSFC of China with Project Nos. 11935007, 11805167.








\begin{thebibliography}{99}

\bibitem{Vitev:2009rd}
  I.~Vitev and B.~W.~Zhang,
  Phys.\ Rev.\ Lett.\  {\bf 104}, 132001 (2010).


\bibitem{Sirunyan:2019dow}
  A.~M.~Sirunyan {\it et al.} [CMS Collaboration],
  Phys.\ Rev.\ Lett.\  {\bf 125}, 102001 (2020)
  [arXiv:1911.01461 [hep-ex]].

\bibitem{Wang:2019xey}
  S.~Wang, W.~Dai, B.~W.~Zhang and E.~Wang,
  EPJC 79,no.9,789 (2019)
  arXiv:1906.01499 [nucl-th].

\bibitem{Wang:2020bqz}
  S.~Wang, W.~Dai, J.~Yan, B.~W.~Zhang and E.~Wang,
  Nucl.Phys. A(2020)
  arXiv:2001.11660 [nucl-th].

\bibitem{Gleisberg:2008ta}
  T.~Gleisberg {\it et al.},
  JHEP {\bf 0902} (2009) 007
  [arXiv:0811.4622 [hep-ph]].


\bibitem{Peterson:1982ak}
  C.~Peterson, D.~Schlatter, I.~Schmitt and P.~M.~Zerwas,
  Phys.\ Rev.\ D {\bf 27}, 105 (1983).

  \bibitem{Cacciari:2011ma}
  M.~Cacciari, G.~P.~Salam and G.~Soyez,
  Eur.\ Phys.\ J.\ C {\bf 72} (2012) 1896

\bibitem{Cao:2013ita}
  S.~Cao, G.~Y.~Qin and S.~A.~Bass,
  Phys.\ Rev.\ C {\bf 88}, 044907 (2013)
  [arXiv:1308.0617 [nucl-th]].

\bibitem{Dai:2018mhw}
  W.~Dai, S.~Wang, S.~L.~Zhang, B.~W.~Zhang and E.~Wang,
  arXiv:1806.06332 [nucl-th];

\bibitem{Wang:2020qwe}
  S.~Wang, W.~Dai, B.~W.~Zhang and E.~Wang,
  arXiv:2005.07018 [hep-ph].

\bibitem{Francis:2015daa}
  A.~Francis {\it et al.}
  Phys.\ Rev.\ D {\bf 92}, no. 11, 116003 (2015)
  [arXiv:1508.04543 [hep-lat]].


\bibitem{Guo:2000nz}
  X.~F.~Guo and X.~N.~Wang,
  Phys.\ Rev.\ Lett.\  {\bf 85} (2000) 3591
  [hep-ph/0005044].

\bibitem{Zhang:2003yn}
  B.~W.~Zhang and X.~N.~Wang,
  Nucl.\ Phys.\ A {\bf 720}, 429 (2003).

\bibitem{Zhang:2003wk}
  B.~W.~Zhang, E.~Wang and X.~N.~Wang,
  Phys.\ Rev.\ Lett.\  {\bf 93} (2004) 072301
  [nucl-th/0309040].

\bibitem{Majumder:2009ge}
  A.~Majumder,
  Phys.\ Rev.\ D {\bf 85} (2012) 014023


\bibitem{Ma:2018swx}
  G.~Y.~Ma, W.~Dai, B.~W.~Zhang and E.~K.~Wang,
  Eur.\ Phys.\ J.\ C {\bf 79} (2019) no.6,  518

\bibitem{Neufeld:2010xi}
  R.~B.~Neufeld,
  Phys.\ Rev.\ D {\bf 83} (2011) 065012

\bibitem{Shen:2014vra}
  C.~Shen, Z.~Qiu, H.~Song, J.~Bernhard, S.~Bass and U.~Heinz,
  Comput.\ Phys.\ Commun.\  {\bf 199} (2016) 61

\bibitem{Wang:2020gxz}
  S.~Wang, W.~Dai, B.~W.~Zhang and E.~Wang,
  in preparation.

\end{thebibliography}
\end{document}